\documentclass[a4paper,fleqn,usenatbib]{mnras}

\usepackage{newtxtext,newtxmath}

\usepackage[T1]{fontenc}
\usepackage{ae,aecompl}

\usepackage{graphicx}	
\usepackage{amsmath}	
\usepackage{amssymb}	
\usepackage{longtable}

\title[SDSS ring galaxy candidates]{Automatic detection of full ring galaxy candidates in SDSS}

\author[Lior Shamir]{
Lior Shamir,$^{1}$\thanks{E-mail: lshamir@mtu.edu}
\\
$^{1}$Kansas State University, Manhattan, KS 65506, USA
}

\date{Accepted XXX. Received YYY; in original form ZZZ}

\pubyear{2019}

\begin{document}
\label{firstpage}
\pagerange{\pageref{firstpage}--\pageref{lastpage}}
\maketitle

\begin{abstract}

A full ring is a form of galaxy morphology that is not associated with a specific stage on the Hubble sequence. Digital sky surveys can collect many millions of galaxy images, and therefore even rare forms of galaxies are expected to be present in relatively large numbers in image databases created by digital sky surveys. Sloan Digital Sky Survey (SDSS) data release (DR) 14 contains $\sim2.6\cdot10^6$ objects with spectra identified as galaxies. The method described in this paper applied automatic detection to identify a set of 443 ring galaxy candidates, 104 of them were already included in the Buta + 17 catalogue of ring galaxies in SDSS, but the majority of the galaxies are not included in previous catalogues. Machine analysis cannot yet match the superior pattern recognition abilities of the human brain, and even a small false positive rate makes automatic analysis impractical when scanning through millions of galaxies. Reducing the false positive rate also increases the true negative rate, and therefore the catalogue of ring galaxy candidates is not exhaustive. However, due to its clear advantage in speed, it can provide a large collection of galaxies that can be used for follow-up observations of objects with ring morphology.

\end{abstract}


\begin{keywords}
Catalogs --- techniques: image processing  --- methods: data analysis --- galaxies: peculiar
\end{keywords}



\section{Introduction}
\label{introduction}


The deployment of autonomous digital sky surveys has enabled the creation of very large databases of galaxy images, and therefore even very rare types of galaxies are assumed to be present in these databases. One of the less common types of galaxies is ring galaxies. Ring galaxies can be separated into several different types \citep{buta1996galactic} such as bar-driven or tidially-driven resonance rings \citep{buta2000resonance}, collisional rings \citep{appleton1996collisional}, polar rings \citep{whitmore1990new,maccio2005origin,reshetnikov1997global,finkelman2012polar,reshetnikov2015polar}, ``Hoag-type'' rings \citep{hoag1950peculiar,brosch85,schweizer1987structure}, and spiral galaxies with ringed bars \citep{buta2001dynamics}. 

Ring galaxies can be classified by their visual morphology into three major sub-classes \citep{theys1976ring}: Empty rings (RE), rings with off-centre nucleolus (RN), and rings with knots or condensations (RK). Another classification scheme for ring galaxies based on their visual appearance separates ring galaxies into ``O-rings'', which have a smooth ring structure and a nucleolus in its centre, and `'P-type'' rings, which have a knotty structure or a nucleolus that is not in the centre of the ring \citep{few1986ring}.


Some ring galaxy catalogues were created using manual analysis of the galaxies in the past six decades. The catalogue of peculiar galaxies of \cite{arp1966atlas} includes two empty ring galaxies, and the \cite{arp1988catalogue} catalogue includes 69 ring galaxies. \cite{struck2010applying} prepared a catalogue of a dozen colliding ring galaxies from SDSS based on reports of volunteers in the Galaxy Zoo on-line forum. The \cite{whitmore1990new} catalogue included 157 polar ring galaxy candidates, and several of these galaxies were confirmed as polar rings \citep{finkelman2012polar}. \cite{madore2009atlas} released an atlas of collisional rings. \cite{garcia2015initiating} identified 16 polar ring galaxy candidates. \cite{buta1995catalog} collected a set of Southern ring galaxies.  \cite{moiseev2011new} and \citep{buta2017galactic} used citizen science annotations and classifications to identify ring galaxy candidates by using the Galaxy Zoo 1 and Galaxy Zoo 2 databases, respectively. These catalogues are efficient in the sense that they have good detection accuracy due to the superior ability of the human brain to analyze galaxy morphology, but because they require very intensive labour, even when using a large number of volunteers it is difficult to perform an exhaustive analysis of the entire image databases collected by modern digital sky surveys. That bandwidth limitation will be magnified when more powerful sky surveys such as LSST see first light. \cite{timmis2017catalog} used computer analysis to release a catalogue of 186 automatically identified ring galaxy candidates in PanSTARRS.

As digital sky surveys become increasingly more powerful, it is clear that manual analysis of the images is not sufficient for comprehensive detection of ring galaxies among millions of galaxy images. That reinforces the use of automation to detect ring galaxies. The ability to identify galaxy morphology automatically can lead to much larger collections of ring galaxies, which can also be useful when more powerful digital sky surveys such as LSST start to collect data.



\section{Galaxy image analysis method}
\label{method}

The data source used in this study is the set of galaxies with spectra in SDSS DR14. SDSS DR14 contains a total of $\sim4.8\cdot10^6$ IDs of objects with spectra, and $\sim2.6\cdot10^6$ of these objects are labeled by SDSS pipeline as galaxies. The mean redshift of these galaxies is 0.38 ($\sigma$=0.24), and the mean g magnitude is $\sim$20.56 ($\sigma$=2.09). The image of each galaxy was obtained by using the {\it cutout} service of SDSS as was done in \citep{kum16}. In summary, the images are downloaded as 120$\times$120 JPG colour images. Since galaxies have different sizes, each galaxy was downloaded several times until 25\% or less of the pixels on the edges of the image have gray value of less than 125. The initial scale was set to 0.25'' per pixel, and it increased by 0.05'' until 25\% or less of the pixels on the edges are not bright, which means that the galaxy fits inside the image \citep{kum16}. 

The JPG images are used because they combine information from the different bands, providing a simple image format that contains  information about the morphology of each object in a manner that is easier to process by machine vision. While the original FITS format allows to make accurate photometric measurements, that accuracy is not required for machine vision systems for the purpose of broad morphological analysis. Therefore, the simple JPG format provides an efficient mechanism for both manual \citep{willett2013galaxy} and automatic \citep{dieleman2015rotation,kum16} analysis. 

Downloading that large dataset of galaxy images required $\sim$16 days. The image analysis method is similar to the method used in \cite{timmis2017catalog}. Each image was converted to a binary map such that all pixels above the threshold were set to 1, and the pixels below the threshold were set to 0. The initial threshold was set to 50, and increased by five until it reached 200.

For each  threshold, the image is inverted, and a 4-connected labeling algorithm is applied to label all objects in the inverted image. If more than one object is detected, it means that the image contained background areas that are inside foreground objects, and therefore could be rings. Since a galaxy can contain many small ares inside the arms, if the size of the background area is less than 10\% of the foreground galaxy the algorithm ignores that background area and does not consider it as a ring candidate. The algorithm is implemented as part of the Ganalyzer galaxy image analysis tool \citep{shamir2011ganalyzer,shamir2011ganalyzer_ascl}.

\section{Ring galaxy candidates}
\label{catalog}

The method described in Section~\ref{method} and also explained in \citep{timmis2017catalog} detected ring galaxy candidates, as listed in Table~\ref{ring_galaxies}. The galaxies are provided with their catalogue number, right ascension and declination of each object.

\begin{figure*}
\centering
\includegraphics[scale=0.80]{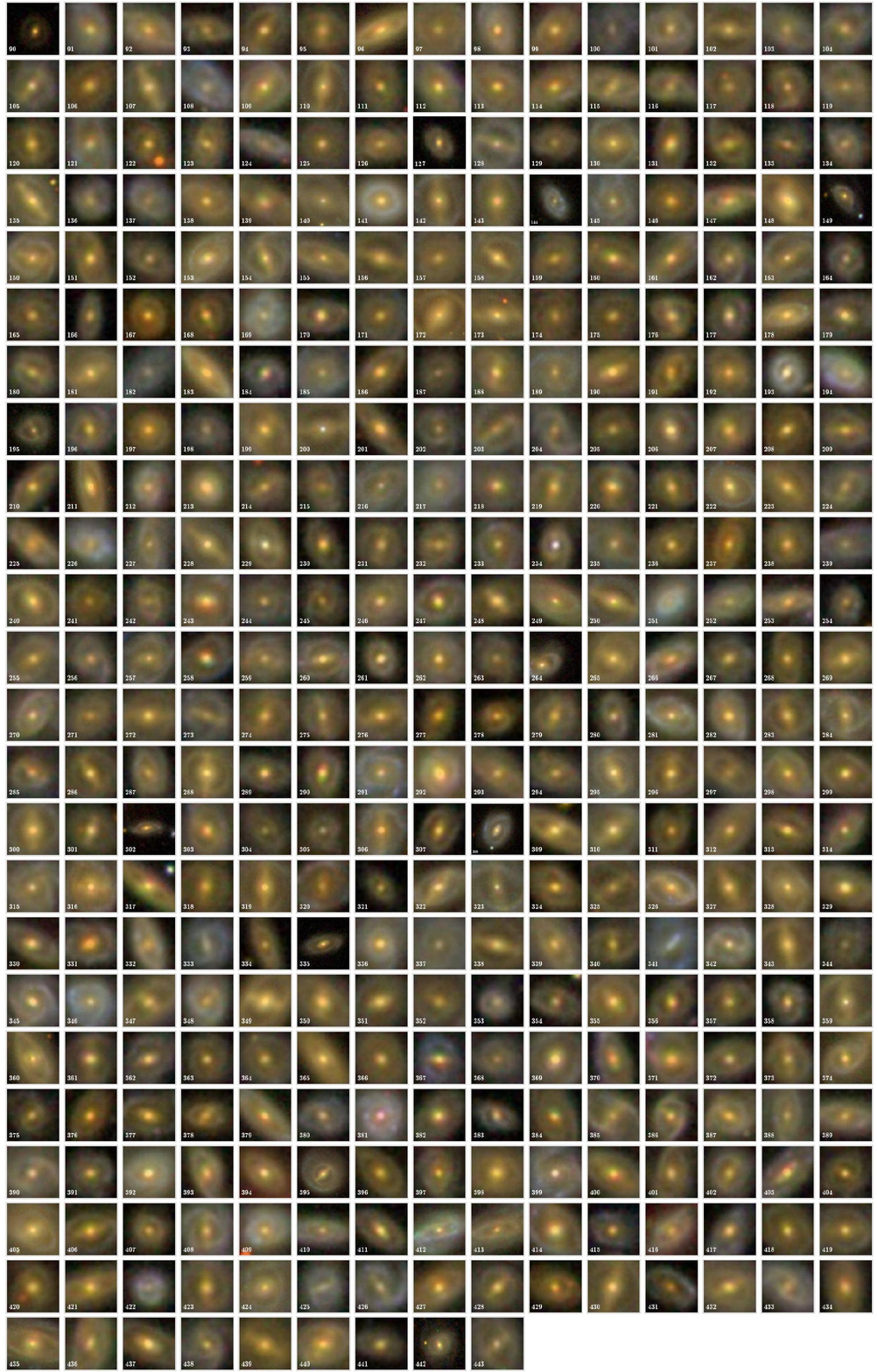}
\caption{SDSS candidates of resonance ring galaxies.}
\label{resonance}
\end{figure*}

\begin{figure*}
\centering
\includegraphics[scale=0.38]{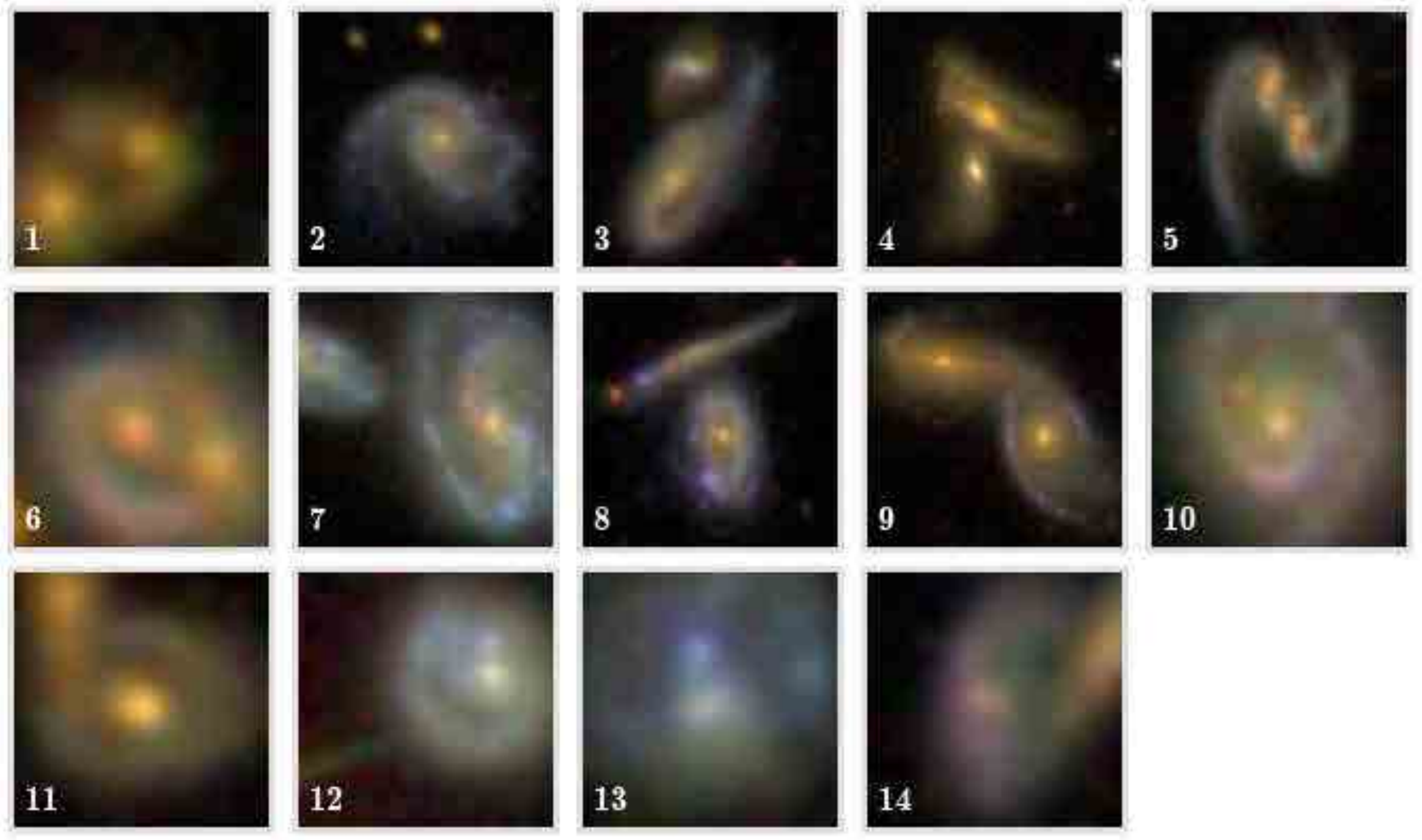}
\caption{SDSS candidates of collisional ring galaxies.}
\label{collisional}
\end{figure*}

\begin{figure*}
\centering
\includegraphics[scale=0.45]{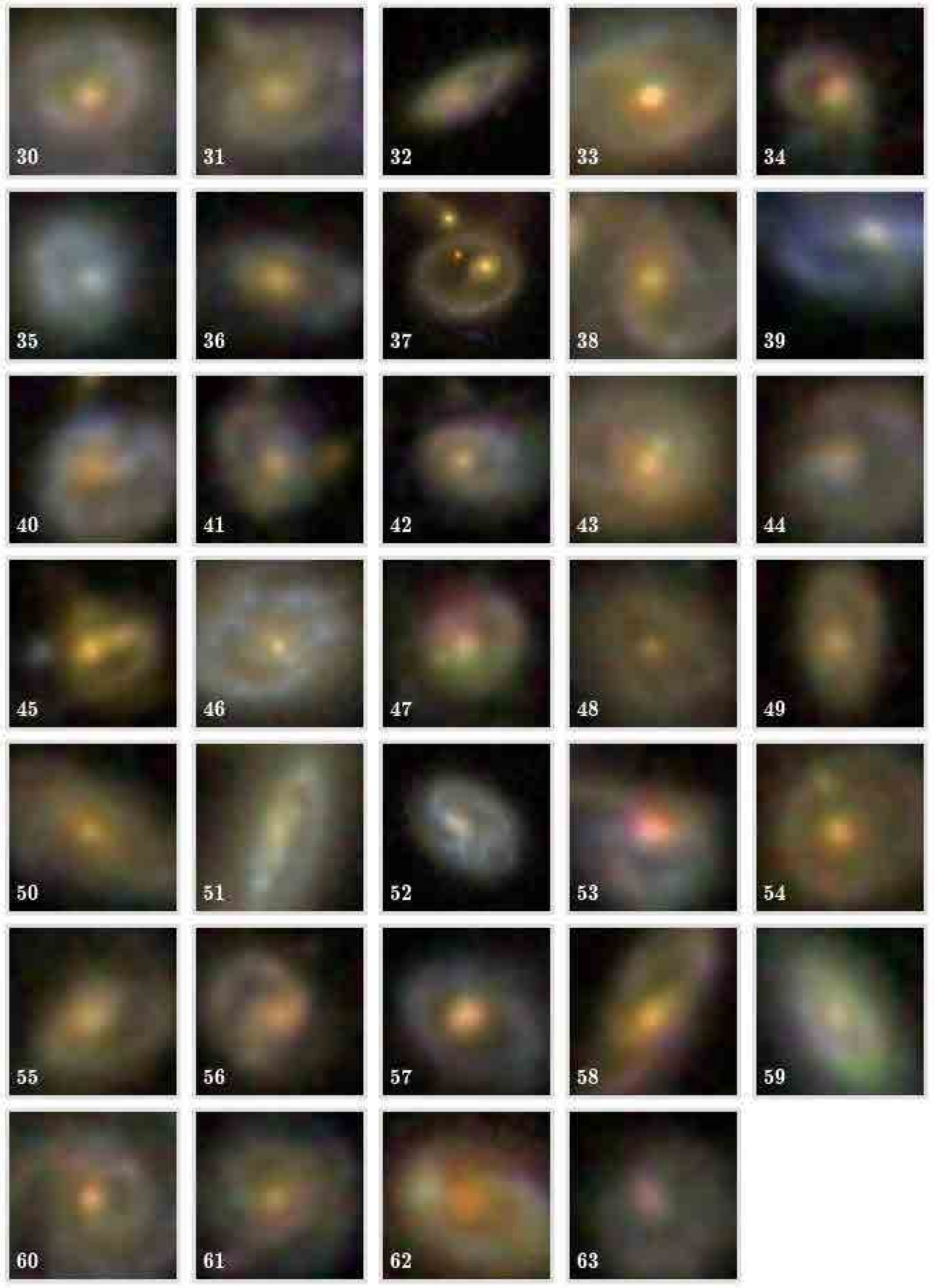}
\caption{SDSS candidates of ring galaxies with off-centre nucleus.}
\label{off_center}
\end{figure*}

\begin{figure*}
\centering
\includegraphics[scale=0.38]{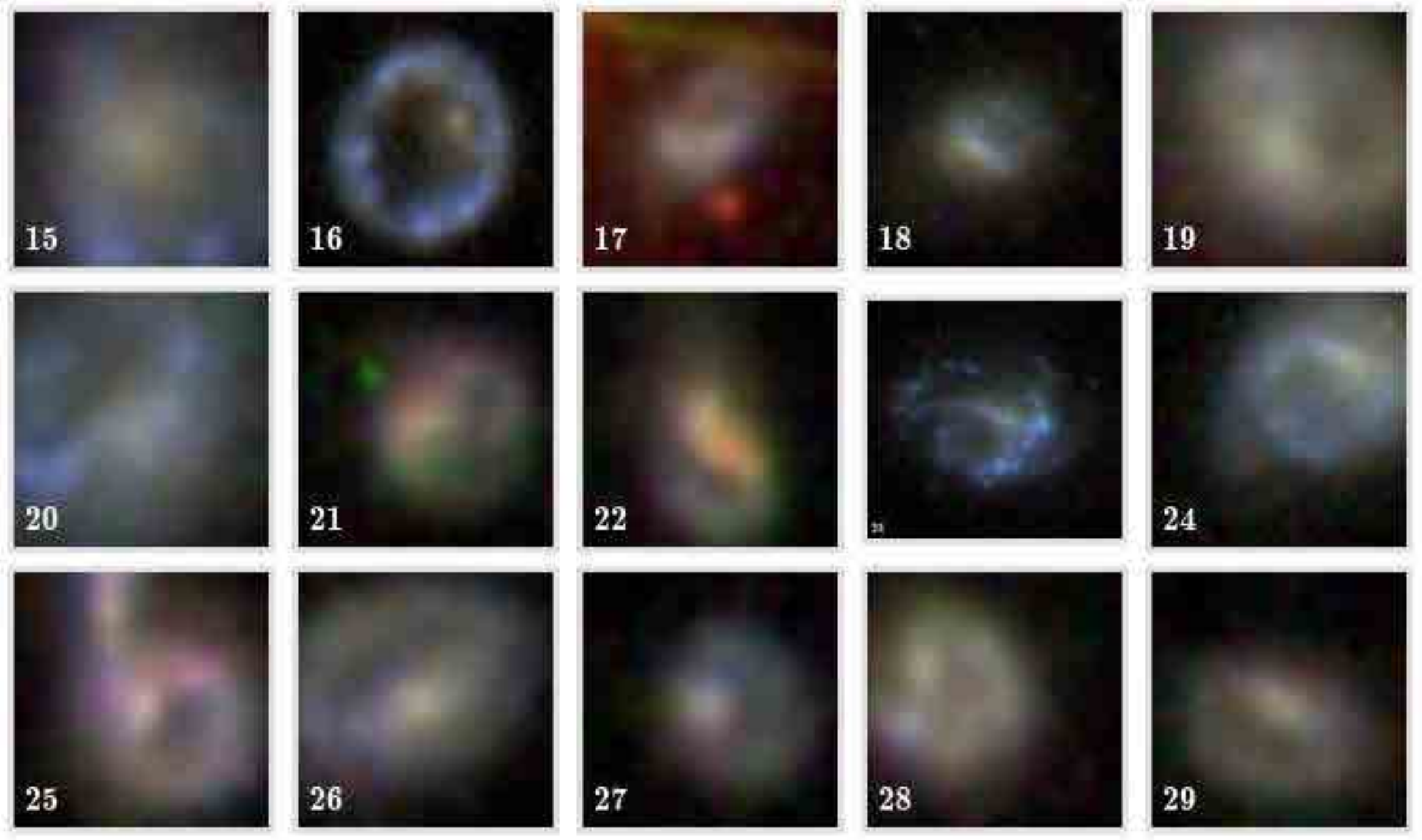}
\caption{SDSS candidates of ring galaxies with no obvious nucleus inside the ring.}
\label{no_nucleus}
\end{figure*}

\begin{figure*}
\centering
\includegraphics[scale=0.38]{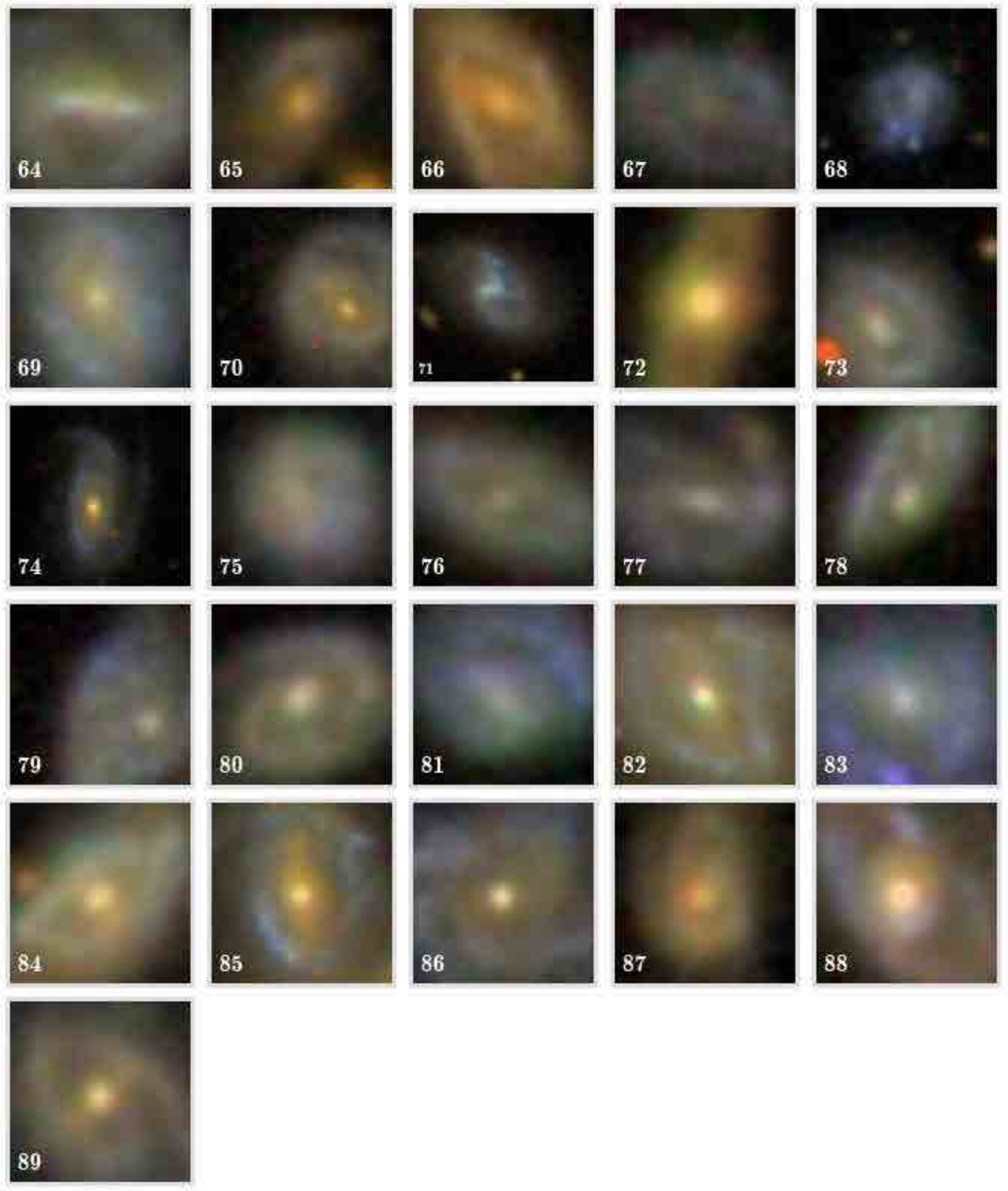}
\caption{Other ring galaxy candidates in SDSS not included in Figures~\ref{resonance},~\ref{collisional},~\ref{no_nucleus}, and~\ref{off_center}.}
\label{others}
\end{figure*}

The images of the galaxies are shown by Figures~\ref{resonance},~\ref{collisional},~\ref{off_center},~\ref{no_nucleus}, and~\ref{others}, showing candidate resonance ring galaxies, collisional rings, rings with an off-centre nucleus, rings with no obvious nucleus, and other rings, respectively.



\scriptsize

\onecolumn


\begin{longtable}[ht]{|l|c|c|c|c|c|l|c|c|c|c|c|l|c|c|c|c|}



\hline
No & RA ($^o$) & Dec ($^o$)  & & No & RA ($^o$) & Dec ($^o$)  & & No & RA ($^o$) & Dec ($^o$)  & & No & RA ($^o$) & Dec ($^o$)  \\
\hline

1 & 249.478 & 45.695 & & 2 & 146.776 & 54.312 & & 3 & 204.487 & -1.718 & & 4 & 134.007 & 42.273 \\ 
5 & 151.691 & 48.628 & & 6 & 174.501 & 48.439 & & 7 & 240.680 & 41.197 & & 8 & 150.674 & 12.624 \\ 
9 & 178.885 & 46.219 & & 10 & 247.738 & 21.791 & & 11 & 242.607 & 17.760 & & 12 & 184.600 & 20.141 \\

13 & 175.241 & 17.219 & & 14 & 177.949 & 2.0095 & & 15 & 151.699 & -0.503 & & 16 & 188.017 & 66.405 \\ 
17 & 313.869 & 0.5356 & & 18 & 147.504 & 47.108 & & 19 & 254.346 & 25.465 & & 20 & 246.945 & 38.943 \\ 
21 & 345.459 & 0.8371 & & 22 & 206.989 & 34.975 & & 23 & 142.281 & 30.142 & & 24 & 212.959 & 31.927 \\ 
25 & 209.728 & 29.576 & & 26 & 121.734 & 13.760 & & 27 & 161.107 & 26.490 & & 28 & 199.832 & 21.625 \\

29 & 200.213 & 12.157 & & 30 & 181.458 & -0.184 & & 31 & 114.872 & 32.724 & & 32 & 130.460 & 45.426 \\ 
33 & 344.860 & 15.151 & & 34 & 32.5462 & 0.8323 & & 35 & 191.767 & 51.582 & & 36 & 169.010 & 52.136 \\ 
37 & 134.635 & 37.087 & & 38 & 196.905 & 49.771 & & 39 & 157.796 & 6.8742 & & 40 & 236.009 & 45.956 \\ 
41 & 212.605 & 57.661 & & 42 & 121.289 & 25.396 & & 43 & 157.020 & 37.820 & & 44 & 202.846 & 10.891 \\ 
45 & 210.915 & 36.730 & & 46 & 226.564 & 10.340 & & 47 & 358.794 & 0.6218 & & 48 & 45.2437 & 0.0599 \\ 
49 & 115.312 & 47.646 & & 50 & 153.830 & 11.924 & & 51 & 182.373 & 39.815 & & 52 & 185.316 & 36.335 \\ 
53 & 212.079 & 29.079 & & 54 & 196.828 & 31.078 & & 55 & 219.011 & 26.653 & & 56 & 232.845 & 20.023 \\ 
57 & 216.339 & 25.179 & & 58 & 246.557 & 13.215 & & 59 & 185.591 & 29.581 & & 60 & 195.454 & 19.008 \\

61 & 125.013 & 11.941 & & 62 & 126.532 & 10.747 & & 63 & 132.706 & 11.309 & & 64 & 21.8792 & 14.819 \\ 
65 & 358.874 & 14.192 & & 66 & 334.402 & 12.700 & & 67 & 343.107 & 13.176 & & 68 & 178.438 & 55.528 \\ 
69 & 136.982 & 42.311 & & 70 & 134.690 & 46.042 & & 71 & 156.889 & 45.897 & & 72 & 224.223 & 49.878 \\ 
73 & 164.944 & 42.657 & & 74 & 164.202 & 44.303 & & 75 & 238.755 & 34.936 & & 76 & 236.775 & 30.955 \\ 
77 & 242.395 & 27.670 & & 78 & 349.965 & 0.4225 & & 79 & 323.714 & 0.3869 & & 80 & 202.922 & 14.237 \\ 
81 & 192.539 & 35.383 & & 82 & 180.130 & 31.947 & & 83 & 161.572 & 29.359 & & 84 & 209.222 & 20.142 \\ 
85 & 174.478 & 21.985 & & 86 & 165.202 & 15.602 & & 87 & 169.606 & 17.286 & & 88 & 21.3182 & -8.873 \\

89 & 179.819 & -1.108 & & 90 & 57.4584 & 0.0522 & & 91 & 211.549 & -1.227 & & 92 & 222.089 & -0.807 \\ 
93 & 173.951 & -0.494 & & 94 & 186.144 & 0.3766 & & 95 & 27.7023 & 13.568 & & 96 & 21.3422 & 14.838 \\ 
97 & 56.1696 & -5.625 & & 98 & 172.324 & -1.708 & & 99 & 188.859 & -3.602 & & 100 & 131.250 & 52.393 \\ 
101 & 121.762 & 45.676 & & 102 & 198.947 & -0.462 & & 103 & 139.699 & 55.705 & & 104 & 129.686 & 50.619 \\ 
105 & 118.790 & 44.173 & & 106 & 120.193 & 47.176 & & 107 & 212.371 & 64.913 & & 108 & 193.725 & 1.5910 \\ 
109 & 232.049 & 2.5300 & & 110 & 216.834 & 1.0258 & & 111 & 200.184 & 1.7388 & & 112 & 161.881 & 2.0791 \\ 
113 & 148.000 & 2.5866 & & 114 & 211.465 & 3.0833 & & 115 & 211.105 & 3.7596 & & 116 & 217.769 & 4.8296 \\ 
117 & 337.496 & -8.593 & & 118 & 326.511 & -7.198 & & 119 & 330.341 & -7.128 & & 120 & 33.5925 & -9.104 \\ 
121 & 12.6542 & -9.068 & & 122 & 327.665 & -8.332 & & 123 & 310.736 & -5.808 & & 124 & 325.317 & -7.257 \\ 
125 & 344.334 & 14.366 & & 126 & 349.625 & 14.826 & & 127 & 3.76 & -10.155 & & 128 & 40.2084 & -7.975 \\ 
129 & 120.524 & 41.188 & & 130 & 136.520 & 51.735 & & 131 & 12.3798 & 15.987 & & 132 & 148.945 & 1.6018 \\ 
133 & 167.917 & 1.5236 & & 134 & 172.665 & 1.5887 & & 135 & 166.954 & 2.3566 & & 136 & 143.045 & 55.219 \\ 
137 & 149.479 & 4.2610 & & 138 & 159.731 & 4.8516 & & 139 & 154.457 & -0.829 & & 140 & 217.412 & 3.2662 \\ 
141 & 216.152 & 4.5591 & & 142 & 142.796 & 52.635 & & 143 & 171.822 & 3.7559 & & 144 & 245.442 & 43.355 \\ 
145 & 198.781 & 62.521 & & 146 & 154.892 & 60.226 & & 147 & 199.446 & 61.082 & & 148 & 227.381 & 54.506 \\ 
149 & 223.241 & 56.502 & & 150 & 235.794 & 1.3288 & & 151 & 236.464 & 1.6150 & & 152 & 254.836 & 32.164 \\ 
153 & 247.893 & 40.565 & & 154 & 247.611 & 41.483 & & 155 & 254.044 & 34.836 & & 156 & 202.138 & -2.215 \\ 
157 & 209.293 & -2.121 & & 158 & 210.519 & -1.357 & & 159 & 203.819 & -2.556 & & 160 & 218.867 & -2.077 \\ 
161 & 217.009 & -1.851 & & 162 & 217.181 & -1.696 & & 163 & 221.867 & -1.633 & & 164 & 258.276 & 33.319 \\ 
165 & 219.552 & -1.517 & & 166 & 214.472 & 6.2075 & & 167 & 331.913 & 11.623 & & 168 & 327.669 & 12.683 \\ 
169 & 344.061 & 12.884 & & 170 & 319.137 & 10.165 & & 171 & 322.616 & 11.734 & & 172 & 313.865 & -1.225 \\ 
173 & 321.890 & -1.188 & & 174 & 337.481 & -0.751 & & 175 & 310.202 & 1.0436 & & 176 & 348.231 & -0.906 \\ 
177 & 22.7615 & 0.6365 & & 178 & 12.0399 & -0.912 & & 179 & 4.22992 & -0.460 & & 180 & 163.053 & 55.220 \\ 
181 & 168.702 & 56.578 & & 182 & 138.148 & 45.262 & & 183 & 177.635 & 55.057 & & 184 & 181.786 & 55.179 \\ 
185 & 185.964 & 56.049 & & 186 & 124.311 & 37.030 & & 187 & 168.055 & 50.536 & & 188 & 121.207 & 29.331 \\ 
189 & 165.645 & 50.582 & & 190 & 120.283 & 29.148 & & 191 & 147.666 & 46.679 & & 192 & 182.677 & 53.037 \\ 
193 & 148.591 & 51.243 & & 194 & 152.806 & 53.516 & & 195 & 141.639 & 48.011 & & 196 & 118.186 & 25.786 \\ 
197 & 164.499 & 51.017 & & 198 & 151.799 & 48.755 & & 199 & 150.764 & 45.597 & & 200 & 115.717 & 22.112 \\ 
201 & 127.439 & 32.611 & & 202 & 138.063 & 39.126 & & 203 & 178.757 & 48.786 & & 204 & 126.713 & 4.4197 \\ 
205 & 149.480 & 55.911 & & 206 & 172.044 & 60.538 & & 207 & 176.594 & 6.8419 & & 208 & 152.865 & 6.6629 \\ 
209 & 170.926 & 7.5019 & & 210 & 170.435 & 9.0033 & & 211 & 172.277 & 8.9885 & & 212 & 195.901 & 8.9922 \\ 
213 & 197.627 & 9.0227 & & 214 & 170.275 & 9.6956 & & 215 & 143.277 & 8.1112 & & 216 & 184.366 & 67.558 \\ 
217 & 183.089 & 68.120 & & 218 & 183.552 & 68.354 & & 219 & 157.983 & 50.684 & & 220 & 158.091 & 50.698 \\ 
221 & 198.592 & 53.077 & & 222 & 189.000 & 54.220 & & 223 & 170.973 & 53.848 & & 224 & 175.943 & 54.442 \\ 
225 & 180.213 & 54.591 & & 226 & 185.335 & 54.761 & & 227 & 242.073 & 38.176 & & 228 & 244.923 & 36.088 \\ 
229 & 245.153 & 36.365 & & 230 & 228.312 & 48.495 & & 231 & 214.146 & 55.481 & & 232 & 204.799 & 57.900 \\ 
233 & 246.322 & 38.792 & & 234 & 235.189 & 47.867 & & 235 & 240.961 & 44.508 & & 236 & 242.529 & 43.458 \\ 
237 & 49.9495 & -0.221 & & 238 & 150.004 & 8.5004 & & 239 & 152.192 & 9.1769 & & 240 & 174.041 & 10.055 \\ 
241 & 159.028 & 45.131 & & 242 & 142.741 & 8.9121 & & 243 & 133.555 & 6.9731 & & 244 & 144.494 & 10.299 \\ 
245 & 124.369 & 7.6804 & & 246 & 143.128 & 11.715 & & 247 & 170.282 & 15.133 & & 248 & 128.269 & 27.860 \\ 
249 & 143.547 & 33.935 & & 250 & 226.157 & 40.372 & & 251 & 157.685 & 40.057 & & 252 & 231.672 & 46.582 \\ 
253 & 242.420 & 39.406 & & 254 & 251.975 & 32.123 & & 255 & 222.494 & 52.609 & & 256 & 211.155 & 54.793 \\ 
257 & 213.933 & 54.043 & & 258 & 211.272 & 43.273 & & 259 & 195.093 & 47.445 & & 260 & 218.954 & 42.539 \\ 
261 & 213.271 & 44.608 & & 262 & 205.493 & 46.874 & & 263 & 163.419 & 42.012 & & 264 & 165.450 & 44.449 \\ 
265 & 190.911 & 44.094 & & 266 & 173.248 & 43.993 & & 267 & 178.491 & 44.535 & & 268 & 209.072 & 12.177 \\ 
269 & 215.581 & 47.935 & & 270 & 204.184 & 51.544 & & 271 & 168.958 & 41.409 & & 272 & 190.235 & 42.905 \\ 
273 & 173.123 & 42.806 & & 274 & 166.485 & 6.3174 & & 275 & 188.748 & 6.6979 & & 276 & 175.650 & 7.0617 \\ 
277 & 186.338 & 42.850 & & 278 & 240.874 & 24.456 & & 279 & 180.837 & 39.608 & & 280 & 238.010 & 26.315 \\ 
281 & 245.992 & 21.820 & & 282 & 193.027 & 39.818 & & 283 & 229.797 & 7.4837 & & 284 & 200.312 & 8.5059 \\ 
285 & 227.996 & 6.0962 & & 286 & 213.297 & 8.4694 & & 287 & 217.921 & 8.0661 & & 288 & 229.609 & 5.2192 \\ 
289 & 230.929 & 5.0401 & & 290 & 227.691 & 6.3661 & & 291 & 193.573 & 51.172 & & 292 & 233.115 & 41.811 \\ 
293 & 238.011 & 39.112 & & 294 & 237.939 & 38.960 & & 295 & 232.652 & 42.717 & & 296 & 201.442 & 40.103 \\ 
297 & 236.668 & 28.128 & & 298 & 214.405 & 38.164 & & 299 & 230.211 & 33.388 & & 300 & 240.475 & 31.892 \\ 
301 & 242.052 & 30.739 & & 302 & 251.416 & 20.549 & & 303 & 250.602 & 26.474 & & 304 & 240.146 & 28.965 \\ 
305 & 241.110 & 7.6260 & & 306 & 243.344 & 7.0388 & & 307 & 225.839 & 11.308 & & 308 & 248.723 & 23.211 \\ 
309 & 211.579 & 36.833 & & 310 & 223.355 & 33.283 & & 311 & 231.612 & 28.338 & & 312 & 232.668 & 27.989 \\ 
313 & 346.065 & 0.6001 & & 314 & 324.529 & -0.638 & & 315 & 344.775 & -0.296 & & 316 & 338.030 & 0.0370 \\ 
317 & 115.757 & 45.120 & & 318 & 124.647 & 54.488 & & 319 & 331.985 & 0.3701 & & 320 & 131.475 & 59.715 \\ 
321 & 20.2657 & -0.300 & & 322 & 29.2562 & -0.278 & & 323 & 23.9027 & 0.0149 & & 324 & 49.2956 & 0.1095 \\ 
325 & 180.414 & 14.055 & & 326 & 203.926 & 13.330 & & 327 & 179.574 & 15.287 & & 328 & 195.905 & 40.248 \\ 
329 & 189.013 & 39.046 & & 330 & 136.555 & 26.672 & & 331 & 124.107 & 20.652 & & 332 & 117.587 & 17.169 \\ 
333 & 221.580 & 31.938 & & 334 & 229.761 & 29.016 & & 335 & 144.877 & 33.526 & & 336 & 156.226 & 35.127 \\ 
337 & 195.074 & 34.944 & & 338 & 204.919 & 33.689 & & 339 & 142.756 & 26.819 & & 340 & 197.703 & 34.078 \\ 
341 & 198.162 & 34.065 & & 342 & 203.709 & 33.309 & & 343 & 209.014 & 32.716 & & 344 & 161.604 & 33.789 \\ 
345 & 197.488 & 30.913 & & 346 & 200.106 & 30.602 & & 347 & 205.873 & 31.003 & & 348 & 203.169 & 31.986 \\ 
349 & 202.401 & 32.400 & & 350 & 196.213 & 31.725 & & 351 & 159.327 & 30.371 & & 352 & 223.132 & 25.337 \\ 
353 & 229.458 & 24.139 & & 354 & 234.594 & 22.445 & & 355 & 233.594 & 23.501 & & 356 & 225.305 & 21.006 \\ 
357 & 234.024 & 18.348 & & 358 & 222.488 & 22.278 & & 359 & 240.338 & 16.306 & & 360 & 204.683 & 26.328 \\ 
361 & 211.438 & 25.392 & & 362 & 228.788 & 21.329 & & 363 & 233.159 & 19.884 & & 364 & 227.019 & 22.308 \\ 
365 & 236.443 & 21.568 & & 366 & 239.405 & 20.755 & & 367 & 236.073 & 16.952 & & 368 & 226.009 & 21.072 \\ 
369 & 229.920 & 19.815 & & 370 & 246.651 & 14.096 & & 371 & 222.187 & 22.990 & & 372 & 208.172 & 23.029 \\ 
373 & 242.006 & 54.611 & & 374 & 137.512 & 22.851 & & 375 & 144.277 & 25.502 & & 376 & 145.949 & 26.374 \\ 
377 & 162.641 & 27.772 & & 378 & 124.292 & 15.915 & & 379 & 166.420 & 29.146 & & 380 & 127.104 & 18.132 \\ 
381 & 131.427 & 19.725 & & 382 & 134.827 & 17.588 & & 383 & 120.268 & 11.429 & & 384 & 177.835 & 26.471 \\ 
385 & 120.352 & 11.916 & & 386 & 139.235 & 19.302 & & 387 & 172.147 & 26.381 & & 388 & 167.698 & 26.375 \\ 
389 & 153.964 & 24.728 & & 390 & 170.867 & 27.510 & & 391 & 176.846 & 28.055 & & 392 & 203.811 & 25.044 \\ 
393 & 162.443 & 22.669 & & 394 & 164.961 & 24.057 & & 395 & 120.475 & 9.6258 & & 396 & 126.934 & 12.233 \\ 
397 & 197.893 & 21.554 & & 398 & 161.396 & 20.692 & & 399 & 198.673 & 21.793 & & 400 & 192.947 & 21.670 \\ 
401 & 233.208 & 15.037 & & 402 & 216.520 & 19.540 & & 403 & 139.723 & 16.857 & & 404 & 186.972 & 19.438 \\ 
405 & 176.196 & 20.125 & & 406 & 204.959 & 18.714 & & 407 & 235.337 & 12.987 & & 408 & 236.543 & 12.982 \\ 
409 & 238.880 & 12.895 & & 410 & 152.186 & 16.807 & & 411 & 177.394 & 18.705 & & 412 & 169.589 & 19.543 \\ 
413 & 176.278 & 19.966 & & 414 & 207.767 & 19.435 & & 415 & 224.104 & 14.542 & & 416 & 235.309 & 16.629 \\ 
417 & 213.323 & 14.342 & & 418 & 208.308 & 16.165 & & 419 & 210.314 & 16.058 & & 420 & 235.333 & 12.337 \\ 
421 & 167.194 & 16.712 & & 422 & 157.998 & 16.320 & & 423 & 177.656 & 17.823 & & 424 & 180.434 & 17.898 \\ 
425 & 163.701 & 16.631 & & 426 & 166.617 & 17.345 & & 427 & 189.966 & 16.422 & & 428 & 183.208 & 18.269 \\ 
429 & 345.502 & 6.2461 & & 430 & 119.332 & 11.206 & & 431 & 26.8014 & -10.21 & & 432 & 128.536 & 11.987 \\ 
433 & 151.758 & 13.983 & & 434 & 130.783 & 11.085 & & 435 & 176.932 & 1.8262 & & 436 & 190.801 & -2.003 \\ 
437 & 190.383 & 1.5136 & & 438 & 257.934 & 64.112 & & 439 & 261.012 & 64.836 & & 440 & 162.739 & 9.2651 \\ 
441 & 222.847 & 57.139 & & 442 & 122.795 & 18.567 & & 443 & 116.312 & 32.762 & &        &             &           \\

\hline
\caption{Ring galaxy candidates identified automatically} \\
\label{ring_galaxies}
\end{longtable}

\normalsize

\twocolumn

\subsection{Comparison of the ring galaxy candidates to previous catalogues}
\label{comparison}

The galaxies in the catalogue were compared to the catalogue of 275 polar ring galaxy candidates in SDSS \citep{moiseev2011new}. That catalogue showed 23 galaxies with a full ring that fit the morphology of the target galaxies shown in Table~\ref{ring_galaxies}. The catalogue IDs of these galaxies are 7, 239, 240, 241, 243, 244, 245, 246, 249, 253, 254, 255, 256, 259, 260, 261, 263, 265, 267, 268, 270, 272, 274. Comparison to the galaxies in Table~\ref{ring_galaxies} shows that none of these galaxies were also included in Table~\ref{ring_galaxies}. Therefore, Table~\ref{ring_galaxies} is clearly not a complete set of all SDSS galaxies with a full rung morphology, and many relevant galaxies with a full ring still exist in the SDSS database. 

Comparing to the ring galaxies identified by \cite{struck2010applying}, one (CGCG 222-022) of the 12 ring galaxies is included in this catalogue. The fact that just one galaxy is included in the catalogue shows that many more ring galaxies still exist in the SDSS database. 

The list of automatically identified galaxies was also compared to the ring galaxies that were identified in SDSS by using citizen science \citep{buta2017galactic}. The \cite{buta2017galactic} catalogue contains 3,962 galaxies that volunteers identified manually by visually inspecting the images through an on-line web-based platform. From the 443 galaxies identified automatically, 104 are included in the \cite{buta2017galactic} catalogue. The careful manual inspection process used in \citep{buta2017galactic} is clearly more accurate than any existing computer algorithm. However, the manual classification and annotation requires substantial labour, and therefore less than $3\cdot10^5$ galaxies were examined. The method described in this paper is automatic, and was applied to a much larger dataset of $\sim2.6\cdot10^6$ galaxies, and therefore includes very many galaxies that were not examined by \cite{buta2017galactic}. 


It can be expected that many of the objects listed in Table~\ref{ring_galaxies} have been identified previously and are part of existing catalogues. Table~\ref{known_galaxies} shows the ring galaxy candidates that were also identified in previous studies. 

\begin{table}
 \centering
  \begin{tabular}{|l|l|l|}
\hline
\# & Identifier & Ring \\
    &               & Reference      \\
\hline

4  & MCG+07-19-002 & \\     
16   &   VII Zw 466 & \cite{zwicky1968catalogue} \\   
23   & MCG+05-23-004 & \\  
37   &  IIHz4 & \cite{zwicky1968catalogue} \\   
39   & MCG+01-27-015 & PGC 31038 \\ 
60 & IC 4074 & \\ 
64     & IC 1706 & \\ 
78  &  NGC 7613 &  \\  
82   & NGC 4031 &  \cite{buta2017galactic} \\ 
85 & NGC 3754 &  \\   
96   & IC 1698 &  PGC 5261 \\   
101    & MCG+08-15-041 & \cite{buta2017galactic} \\   
110    & IC 1010 &  \\  
130   & NGC 2740 &  \cite{buta2017galactic} \\  
140   & NGC 5636 &  PGC 51785 \\
141   & IC 1007 &  PGC 51465 \\
145    & MCG+11-16-015 & \cite{buta2017galactic} \\  
148    & NGC 5876 &  \cite{buta2017galactic} \\  
153  & NGC 6184 &  \\  
154  &  UGC 10430 &  PGC 58385 \\
155 & UGC 10615 &  \\   
181  & MCG+10-16-093 &  \\ 
183  & MCG+09-19-213 & \cite{buta2017galactic} \\  
189  & UGC 6109 &  \\  
199  & MCG+08-18-057 &  PGC 29124 \\ 
203   & MCG+08-22-038 &  \\ 
211   & IC 699 & \cite{buta2017galactic} \\  
216  & MCG+11-15-044 &  \\ 
222    & NGC 4566 &  \\  
225   & MCG+09-20-062 &  \\ 
228      & UGC 10342 &  \cite{buta2017galactic} \\  
240  &  IC 2941 &  \cite{buta2017galactic}  \\  
250    & UGC 9691 &  \\  
265 &  MCG+07-26-043 &  \cite{buta2017galactic} \\
282 &  IC 3844  &  \cite{buta2017galactic}  \\   
292   &   CGCG 222-022 & \cite{struck2010applying} \\  
295   & NGC 5947 &  \\ 
308      & MCG+04-39-016 &  \\  
316  &  UGC 12068 &  PGC 69089 \\
319 & MCG+00-56-009 &  \\ 
323   & MCG+00-05-013 & PGC 5928 \\ 
326   & IC 901 &  \\ 
328     & IC 4135 &  \\ 
337  & MCG+06-29-011 &  \\ 
341   & MCG+06-29-059 &  \\ 
349  &  UGC 8484 &  PGC 47369 \\ 
359    & UGC 10134 &  \\
360 &  MCG+05-32-048  &  \\ 
374   &  IC 2441C &   \\  
395 & MCG+02-21-005 & \cite{buta2017galactic} \\  
405  & UGC 6719 &  \\  
413   & MCG+03-30-094 &  \\  
424     & MCG+03-31-015 &  \\ 
435   & UGC 6769 & \cite{buta2017galactic} \\   
440  &  NGC 3429 &  \cite{buta2017galactic} \\ 

   \hline
  \end{tabular}
\caption{Galaxies that are part of previous catalogues}
\label{known_galaxies}
\end{table}

\subsection{Distribution and photometry of the ring galaxy candidates}
\label{distribution}

\begin{figure}
\centering
\includegraphics[scale=0.72]{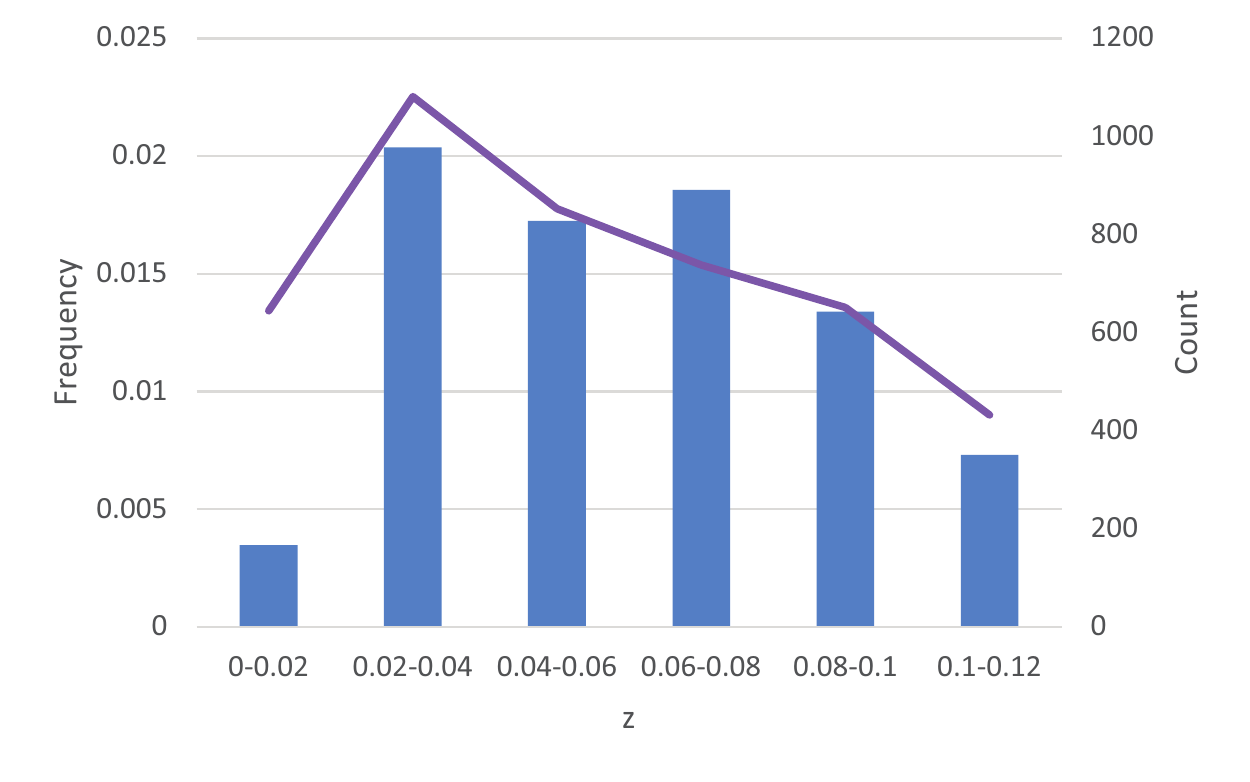}
\caption{The number and frequency of the ring galaxy candidates in Table~\ref{ring_galaxies} combined with the galaxies of the Buta (2017) catalogue. The line shows the number of galaxies in each redshift range, and the bars show the frequency in the entire galaxy population.}
\label{z_distribution}
\end {figure}

As mentioned in Section~\ref{method}, the galaxies in the catalogue described in Section~\ref{catalog} are galaxies detected among the subset of SDSS DR14 galaxies that have spectra. The galaxies included in the \cite{buta2017galactic} catalogue are also galaxies with nuclear spectra. Because the galaxies are galaxies with spectra, their distribution in the sky is not uniform, but a distribution that corresponds to the spectroscopy survey of SDSS DR14. Therefore, the majority of the ring galaxy candidates are in the RA range of 120$^o$-240$^o$. Figure~\ref{z_distribution} shows the distribution of the galaxies in Table~\ref{ring_galaxies} combined with the galaxies of the \cite{buta2017galactic} catalogue by their redshift. The figure shows the number of galaxies, as well as their frequency among the galaxies with spectra in DR14 in the same redshift range. The Petrosian radius of all galaxies is larger than 5.5'', which is large enough to allow the identification of the galaxy morphology \citep{timmis2017catalog}.

\begin{figure}
\centering
\includegraphics[scale=0.62]{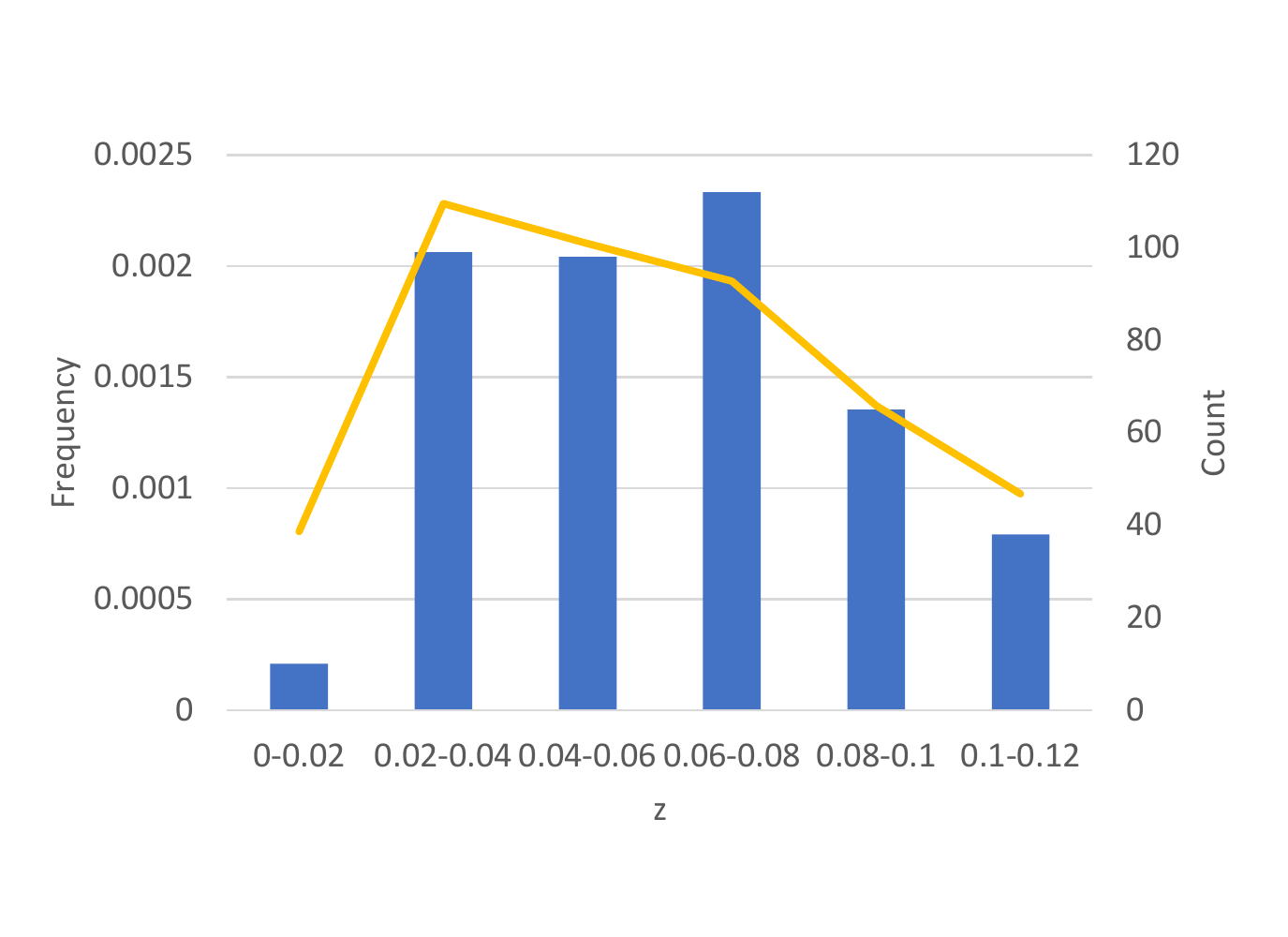}
\caption{The number (line) and frequency (bars) of the ring galaxy candidates in Table~\ref{ring_galaxies} by redshift range.}
\label{z_distribution_old}
\end {figure}

The graph shows that the frequency of the ring galaxy candidates in the catalogue starts to decline when the redshift is higher than 0.08. That can be explained by the less detailed morphology of the imaged galaxies when the redshift gets higher, which does not allow clear identification of morphological details such as the presence of a full ring. The low frequency in the 0-0.02 range can be explained by a higher number of objects misidentified as galaxies by the SDSS pipeline, but are in fact not extra-galactic objects. In any case, in the redshift range of 0-0.02 the number of detected ring galaxy candidates is very small, and does not allow meaningful statistical analysis. Figure~\ref{z_distribution_old} shows the distribution of the ring galaxy candidates in Table~\ref{z_distribution_old}. As the graph shows, the distribution is not substantially different from the distribution in the \cite{buta2017galactic} catalogue.

\begin{figure*}
\centering
\includegraphics[scale=0.70]{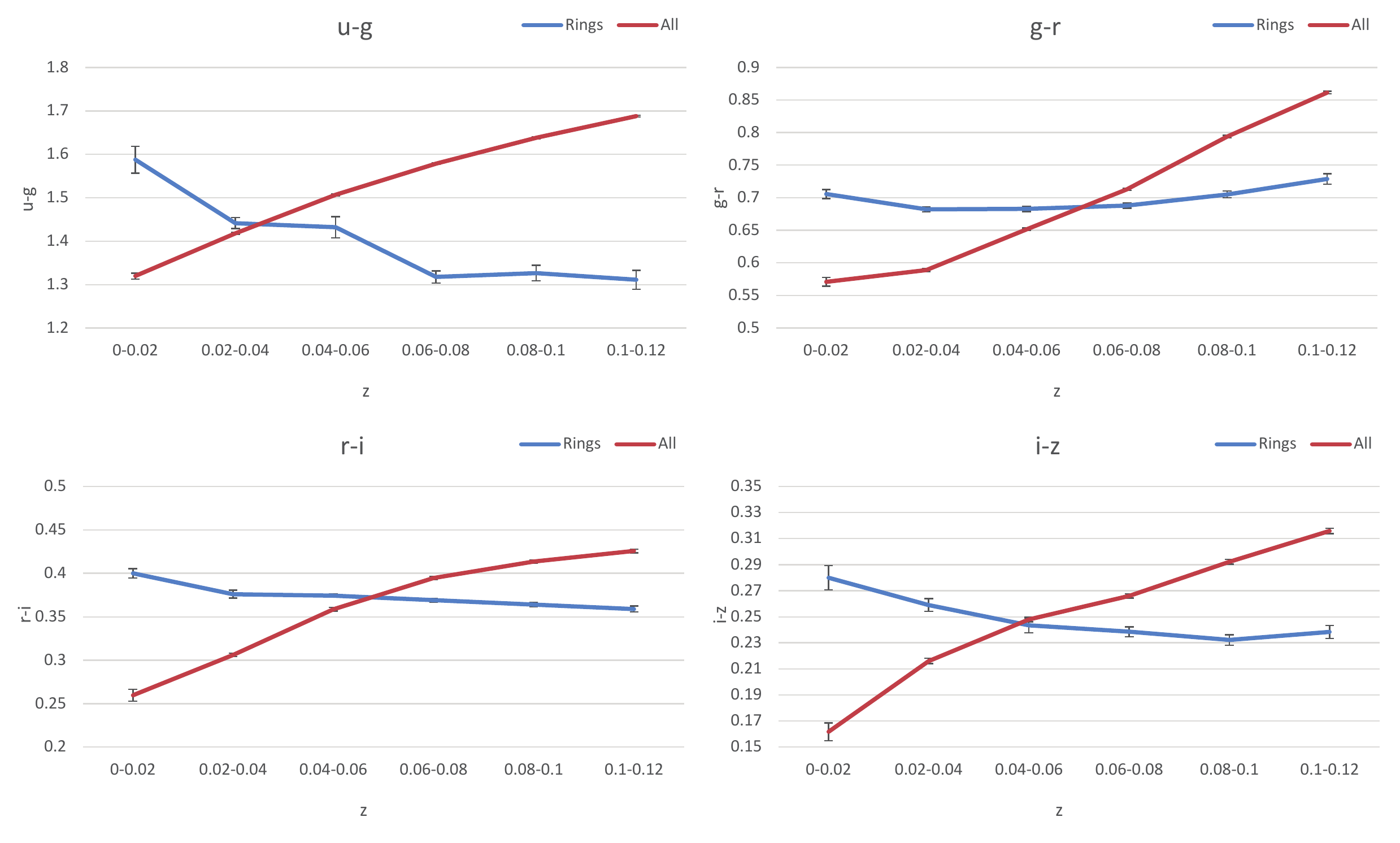}
\caption{Colour differences between ring galaxy candidates and all galaxies.}
\label{color}
\end {figure*}

Figure~\ref{color} shows the colour differences between the ring galaxy candidates including the galaxies of the \cite{buta2017galactic} catalogue, and all other DR14 galaxies with spectra and Petrosian radius larger than 5.5''. The graph shows that the u-g, r-i, and i-z declined with the increase in redshift for the galaxies identified as ring galaxies. That decline is in opposite trend to the other galaxies with spectra and Petrosian radius larger than 5.5''. Also, the colour of the ring galaxy candidates changed in a more moderate manner with the redshift compared to the general galaxy population in SDSS DR14. That can be explained by the more morphologically homogeneous population in the set ring galaxy candidates, compared to the population of galaxies in SDSS. It should be noted that the majority of ring galaxy candidates are selected from the Galaxy Zoo 2 dataset, which are not a random selection of galaxies.

\begin{figure*}
\centering
\includegraphics[scale=0.70]{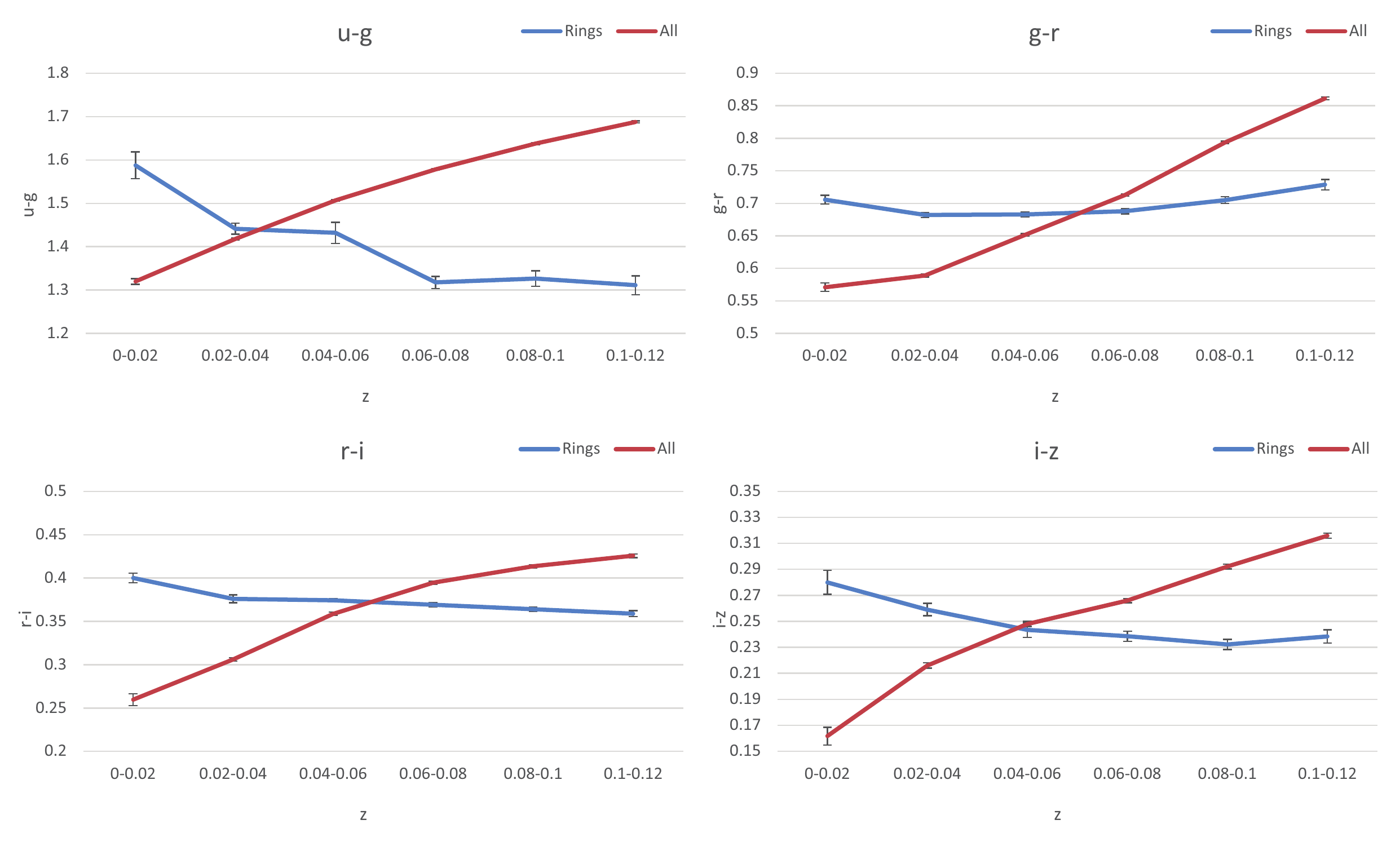}
\caption{Colour differences between ring galaxy candidates of Table~\ref{ring_galaxies} and all galaxies with Petrosian radius larger than 5.5''.}
\label{color_old}
\end {figure*}

Figure~\ref{color_old} shows the colour differences between the ring galaxies in Table~\ref{ring_galaxies} and the other SDSS DR14 galaxies with spectra, and Petrosian radius larger than 5.5''. The graphs show no significant differences between the colour of the ring galaxy candidates and the colour of other galaxies, with the exception of the u-g colour. The u-g of the ring galaxy candidates is lower in all redshift ranges compared to the u-g colour of the other galaxies in SDSS DR14 that have Petrosian radius larger than 5.5''. The mean u-g of the ring galaxy candidates is 1.482$\pm$0.018, while the mean u-g of all other DR14 galaxies with Petrosian radius larger than 5.5'' is 1.572$\pm$0.0007, and therefore the difference is statistically significant ($P<0.001$). The difference in the blue colour can be explained by the fact that rings in star-forming galaxies have a larger visible contract, and therefore can be detected more easily in distant galaxies compared to the redder rings in the same redshift ranges. That can therefore increase the number of blue galaxies among ring galaxies compared to the general galaxy population.

The graphs also show substantial difference in all colours for galaxies in the redshift range of 0-0.02. That can be explained by stars identified by error as galaxies in the SDSS photometric pipeline. However, due to the small number of ring galaxies in that range no meaningful statistical analysis of the difference is possible. It should be noted that the galaxies in the \cite{buta2017galactic} catalogue are bright and large objects selected by Galaxy Zoo 2, and are much larger than the objects in Table~\ref{ring_galaxies}. The mean Petrosian radius  (r band) of the galaxies in the \citep{buta2017galactic} is $\sim$19.28'', while it is $\sim$9.67'' for the galaxies in Table~\ref{ring_galaxies}.

\section{Limitations of the method}
\label{limitations}

The method used in this study aims at analyzing very large databases of galaxy images that might be too big to analyze manually, even when using crowdsourcing. That can only be done by automation. However, due to the very large databases of galaxy images, even a small false positive rate can lead to a very large number of false positive instances that becomes very difficult to handle manually. For instance, in the database used in this paper an algorithm with detection accuracy of 99\% (which is normally considered extremely high in machine vision standards), would generate a dataset of $\sim2.6\cdot10^5$ false positives. Therefore, a practical application of the method requires to minimize the false positive rates. Since machine vision clearly does not meet the accuracy level of the human brain, achieving a low false positive rates require the sacrifice of some of the true positives. 

\begin {figure}[h]
\centering
\includegraphics[scale=0.60]{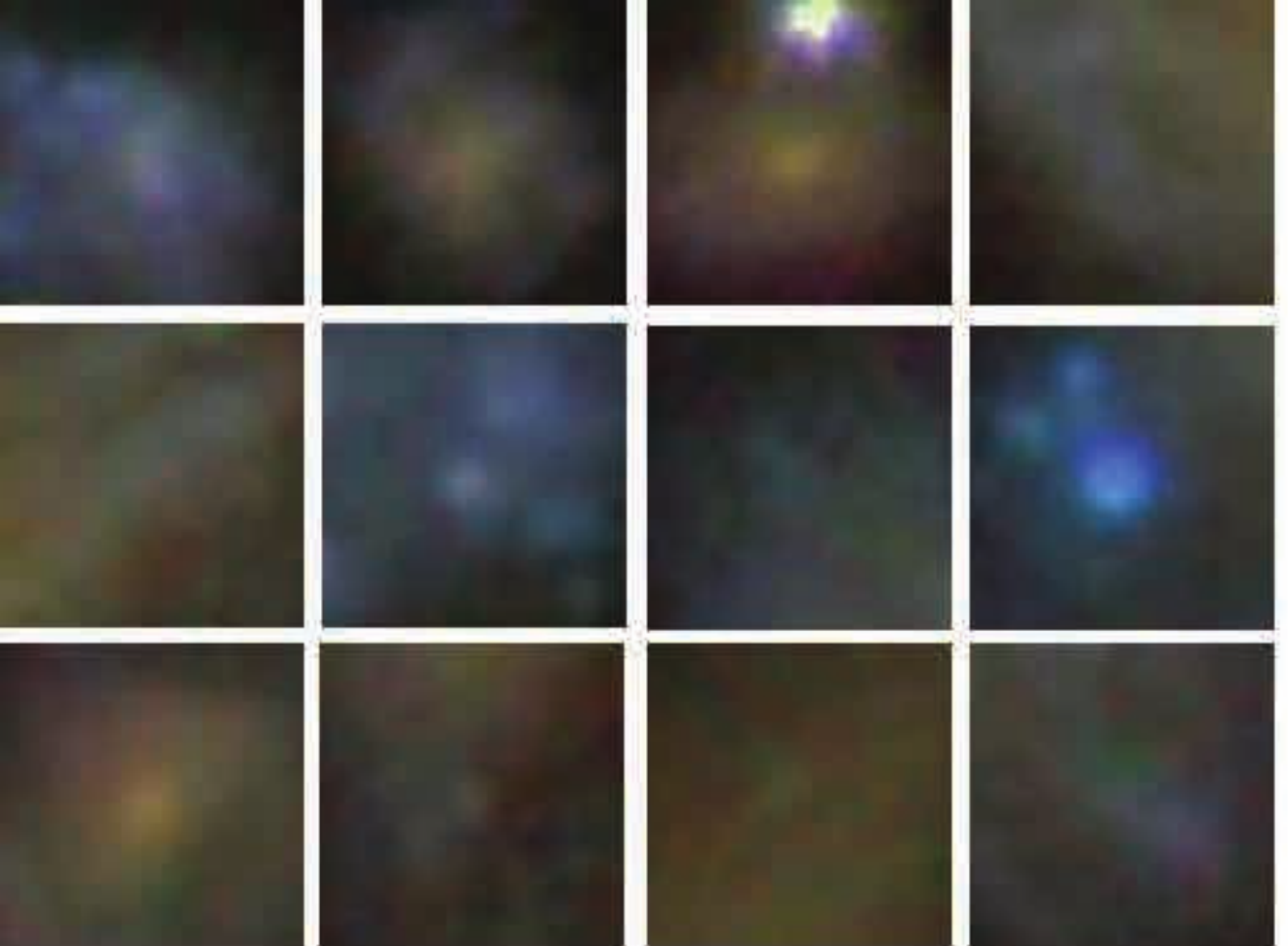}
\caption{Example of galaxies that would have been flagged as ring galaxies below a graylevel threshold of 50.}
\label{dim_rings}
\end {figure}

As mentioned in Section~\ref{method}, pixels below the threshold level of 50 were considered not sufficiently bright and were ignored. The JPEG threshold of 50 
is in some cases high, and can lead to the exclusion of many ring galaxies such as the Hoag object \citep{schweizer1987structure}, which has a clear but relatively dim ring compared to some other ring galaxies. However, lowering the threshold leads to a high number of false positives. For instance, Figure~\ref{dim_rings} shows examples of objects that are not ring galaxies, but the algorithm would have flagged them as rings if a lower graylevel threshold would have been applied. Such objects are very common in the SDSS dataset, and many of them are flagged as galaxies by the SDSS photometric analysis pipeline. Since the method is designed to work with very large databases without the use of manual analysis, it sacrifices the detection of true positives, as even a small rate of false positives leads to an unmanageable output that requires a substantial step of manual analysis.

\begin {figure*}[h]
\centering
\includegraphics[scale=0.60]{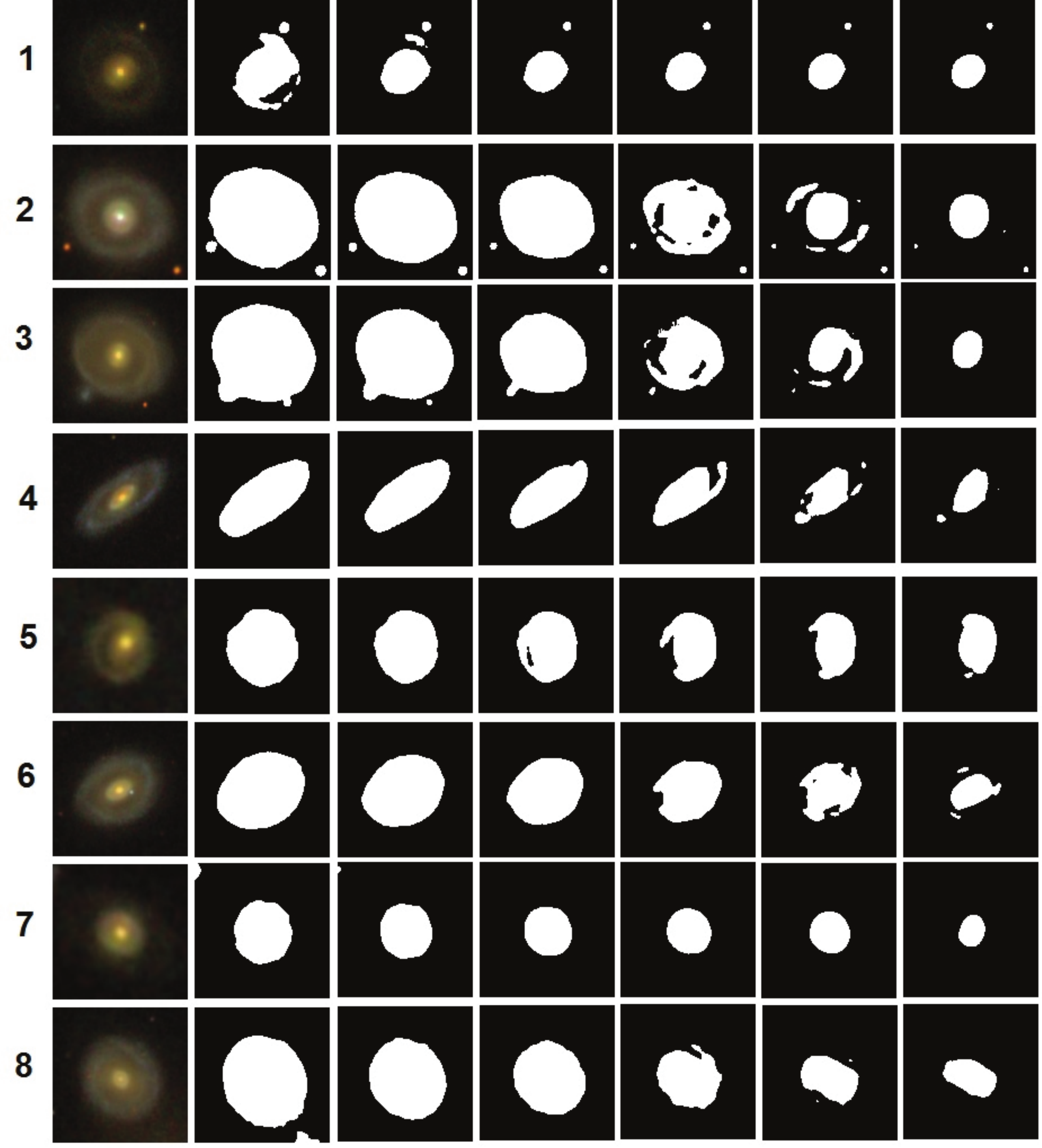}
\caption{Ring galaxies from Struck (2010) catalogue that were not detected by the method, and the binary transformation with different thresholds.}
\label{undetected}
\end {figure*}

As discussed above, avoiding false positives is an important requirement, since due to the very large size of the database even a small false positive rate can make the method unusable. As a result, the detection method also has a high true negative rate, and many ring galaxies might not be detected by the method. To test the behavior of the methods and characterize the ring galaxies that it might fail to detect, ring galaxies from \citep{struck2010applying} that were not detected by the method were examined. Figure~\ref{undetected} shows the first galaxies from the \citep{struck2010applying} Galaxy Zoo sample that were not detected by the method. The figure also shows the binary transformation of each image with different threshold levels.

For galaxies 2, 3, 5, and 6, a small background area surrounded by foreground pixels can be seen. However, these areas are smaller than 10\% of the foreground, and therefore these galaxies are not flagged as ring candidates. In galaxies 1, 4, 6, and 8 part of the ring can be seen in the binary transform, but in none of the threshold levels the ring is complete in the sense that the background is completely surrounded by foreground pixels. For instance, in galaxy 4 the ring opens in the top right part of the galaxy. That happens because the ring is dimmer in that part, and the pixels in that part of the ring do not pass the threshold of the rest of the ring. In ring 6, the lower left part of the ring is dimmer than the rest of the ring, and therefore the ring cannot be detected by the method. In galaxy 3 the luminosity of the area inside the ring is not consistent, and therefore the ring is connected to the nucleus of the galaxy in the binary mask of the image. The same can also be seen in galaxy 5 and galaxy 2.

In galaxy 7, the ring is made of a slightly bluer colour, but the pixel intensity of the ring is not higher than the intensity of the pixels between the ring and the nucleus. Since the method first converts the pixels to grayscale, rings that are visible because they have different colour than the rest of the galaxy will not be detected.

These examples show that the method is mostly dependent on the consistency of the luminosity of the ring, as well as the part of the galaxy inside the ring. Rings that their luminosity varies might not be detected by the method because parts of the ring might not pass the luminosity threshold, leaving the ring in the binary mask open. The same is also for variation inside the ring. If the luminosity inside the ring varies, some parts inside the ring might pass the luminosity threshold and prevent the detection of the ring. Therefore, the method will not always detect ring galaxies that the luminosity of the ring or the parts inside it is not consistent across the different areas.

\section{Conclusion}
\label{conclusion}

While ring galaxies are relatively rare, it can be assumed that the number of ring galaxies within a certain set of galaxies increases with the size of the dataset. The application of automatic identification can therefore allow the detection of such galaxies in very large databases, and is not limited by the availability of human resources that can scan the database manually. When much larger databases such as then Large Synoptic Survey Telescope (LSST) are collected, automatic detection will be able to identify many more ring galaxies.

The collection of ring galaxy candidates described in this paper is clearly not exhaustive, as evident by the differences between the galaxies in this catalogue and the galaxies in the catalogues of \cite{moiseev2011new}, \cite{buta2017galactic}, or \cite{struck2010applying}. Automatic analysis is still not as accurate as the human eye and brain, especially for the non-trivial problem of galaxy image analysis. However, automatic analysis has the clear advantage of analyzing data much faster than any human or group of humans. The purpose of the approach described in this paper is to analyze very large databases of galaxies, under the assumption that even a small true positive rate can lead to large catalogues of ring galaxies. 

Due to its higher sensitivity, the \cite{buta2017galactic} catalogue of manually classified  ring galaxies in SDSS already contains 104 of the galaxies identified in this study. But because the computer analysis method can scan much more galaxies with no cost of human labour, the vast majority of the galaxies identified in this study are not included in previous catalogues. It should be mentioned that the set of galaxies with spectra used as the initial database is not a completely random subset of SDSS galaxies, but selected by a certain algorithm \citep{reid2015sdss}. 

While manual analysis of galaxy morphology has provided good collections of ring galaxies, the labour-intensive efforts required to compile such catalogues reduce the total number of galaxies that can be analyzed. As digital sky surveys are becoming increasingly more powerful, it is clear that automation will be required to analyze these databases and turn them into data products that enable scientific discoveries.

\section{Acknowledgments}
This research was supported by NSF grants AST-1903823 and IIS-1546079. I would like to thank the anonymous reviewer for the insightful comments that helped to improve the paper. Funding for the SDSS and SDSS-II has been provided by the Alfred P. Sloan Foundation, the Participating Institutions, the National Science Foundation, the US Department of Energy, the National Aeronautics and Space Administration, the Japanese Monbukagakusho, the Max Planck Society, and the Higher Education Funding Council for England. The SDSS Web Site is http://www.sdss.org/. The SDSS is managed by the Astrophysical Research Consortium for the Participating Institutions. The Participating Institutions are the American Museum of Natural History, Astrophysical Institute Potsdam, University of Basel, University of Cambridge, Case Western Reserve University, University of Chicago, Drexel University, Fermilab, the Institute for Advanced Study, the Japan Participation Group, Johns Hopkins University, the Joint Institute for
Nuclear Astrophysics, the Kavli Institute for Particle Astrophysics and Cosmology, the Korean Scientist Group, the Chinese Academy of Sciences (LAMOST), Los Alamos National Laboratory, the Max Planck Institute for Astronomy (MPIA), the Max Planck Institute for Astrophysics (MPA), New Mexico State University, Ohio State University, University of Pittsburgh, University of Portsmouth, Princeton University, the United States Naval Observatory and the University of Washington.

\bibliographystyle{mnras}
\bibliography{ms}

\bsp	
\label{lastpage}
\end{document}